\documentclass [12pt]{article}

\usepackage {graphicx}
\usepackage {longtable}
\begin{document}

\begin{center}
\bf{\LARGE Equation of State for Nuclear Matter in Relativistic
Mean-Field Theory and Maxwellian Phase Transition to Strange Quark
Matter}\footnote{Presented at the 8th Conference Quark Confinement
and the Hadron Spectrum, Sept. 1-6, 2008, Mainz, Germany}

\vskip10pt{\bf{Grigor Alaverdyan }
\vskip5pt{\textit {\small
Yerevan State University, A.Manoogyan str. 1, 0025 Yerevan,
Armenia
\\ E-mail: galaverdyan@ysu.am}}}
\end{center}

\begin{quote}
{\small \textbf{Abstract.} Equation of state for superdense
nuclear matter is considered in the framework of relativistic
mean-field theory, when also the scalar-isovector $\delta$-meson
effective field is taken into account. Assuming that the
transition to the strange quark matter is a usual first-order
phase transition described by Maxwells construction, the changes
of the parameters of phase transition caused by the presence of
$\delta$- meson field are investigated. To describe a quark phase
the advanced version of the MIT bag model is used, in which the
interactions between $u, d, s$ quarks are taken into account in
the one-gluon exchange approximation. For different values of the
bag constant $B$, some series of the equations of state of matter
with deconfinement phase transition are constructed. Also the
upper bound, $B_{cr}$, corresponding to the unstable state of the
infinitizimal quark core in a neutron star is found. \vskip5pt

 }
\end{quote}

\section{Introduction}

Within the framework of Quantum Hadro-Dynamics (QHD) the quantum-theoretical
approach allows to adequately enough describe properties both of nuclear
matter and finite nuclei, considering them as a system of strongly
interacting baryons and mesons. One of the similar theories is the
relativistic mean-field theory (RMF) which is effectively used [1].
Inclusion of the scalar-isovector $\delta$-meson in this scheme and
investigation of its influence on low density asymmetric nuclear matter
properties was realized in Ref.[2-3]. In this paper the EOS of superdense
nuclear matter is studied in the framework of RMF approach, the
deconfinement phase transition from hadronic matter to quark matter is
analyzed, and the role of scalar-isovector $\delta$-meson in this process is
investigated.

\section{Deconfinement phase transition parameters}

The relativistic Lagrangian density of an interacting many-particle system
consisting of nucleons and $ \sigma,~\omega,~\rho,~\delta$ -mesons in QHD is
given by
\begin{eqnarray}
\mathcal{L}_{\sigma\omega\rho\delta} (\sigma(x),\omega _{_{\mu }}(x),\vec{%
\rho }_{_{\mu }}(x), \vec{\delta}(x)) = \mathcal{L}_{\sigma\omega
\rho}(\sigma(x),\omega _{_{\mu }}(x),\vec{\rho }_{_{\mu }}(x))-U(\sigma(x))+
\mathcal{L}_{\delta}(\vec{\delta}(x)),
\end{eqnarray}%
where $\mathcal{L_{\sigma\omega \rho}}$ is the linear part of relativistic
Lagrangian density without $\delta$-meson field [4],

$U(\sigma)=\frac{b}{3}m_{_{N}}\left( g_{_{\sigma }}\sigma \right) ^{3}+\frac{%
c}{4}\left( g_{_{\sigma }}\sigma \right) ^{4}$ and $\mathcal{L_{\delta}}$$(%
\vec{\delta})=g_{_{\delta }} \bar {\psi}_{N} \vec{\tau }_{N} \vec{\delta }%
\psi_ {N}+\frac{1}{2}\left(\partial _{_{\mu }}\vec{\delta}\partial ^{^{\mu
}} \vec{\delta}-m_{_{\delta}}\vec{\delta}^{2}\right)$ are the $\sigma$-meson
self-interaction term and contribution of the $\delta$-meson field,
respectively. This Lagrangian density (2.1) contains the meson-nucleon
coupling constants, $g_{_{\sigma }},~ g_{_{\omega }},~g_{_{\rho
}},~g_{_{\delta}}$ and also parameters of $\sigma$-field self-interacting
terms, $b$ and $c$. Constants of RMF theory, $a_{i}=\left(g_{i}/m_{i}%
\right)^2$ $(i=\sigma,~ \omega,~ \rho,~ \delta)$ and $b$, $c$ are
numerically determined to reproduce such empirically known characteristics
of symmetric nuclear matter at saturation density as the mass of bare
nucleons, $m_{N}=938.93~MeV$, the nucleon effective mass factor, $%
\gamma=m_{N}^{\ast}/m_{N}=0.78$, the baryon number density at saturation, $%
n_{0}=0.153~fm^{-3}$, the binding energy per baryon, $f_{0}=-16.3~MeV$, the
incompressibility modulus, $K=300~MeV$, and asymmetry energy $%
E_{sym}^{(0)}=32.5~MeV$.

In this paper $a_{\delta}=2.5~fm^2$ is chosen, according to Ref.[2]. The
values of obtained RMF theory parameters are listed in Table 1.

\begin{table}[h]
\caption{The constants of RMF theory with ($\protect\sigma \protect\omega
\protect\rho \protect\delta$) and without ($\protect\sigma \protect\omega
\protect\rho$) $\protect\delta$-meson field.}
\label{1}\centering
\begin{tabular}{|c|c|c|c|c|c|c|}
\hline
& $a_{\sigma },\ fm^{2}$ & $a_{\omega },\ fm^{2}$ & $a_{\delta },\ fm^{2}$ &
$a_{\rho },\ fm^{2}$ & $b,\ fm^{-1}$ & $c$ \\ \hline
$\sigma \omega \rho $ & $9.154$ & $4.828$ & $0$ & $4.794$ & $1.654\cdot
10^{-2}$ & $1.319\cdot 10^{-2}$ \\ \hline
$\sigma \omega \rho \delta $ & $9.154$ & $4.828$ & $2.5$ & $13.621$ & $%
1.654\cdot 10^{-2}$ & $1.319\cdot 10^{-2}$ \\ \hline
\end{tabular}%
\end{table}
The results of our analysis show that the scalar - isovector $\delta$-meson
field inclusion increases the value of the energy per nucleon, besides, this
change is strengthened with the increase of the nuclear matter asymmetry
parameter, $\alpha=(n_{n}-n_{p})/n$. The $\delta$-field inclusion leads to
the increase of the EOS stiffness of nuclear matter due to the splitting of
proton and neutron effective masses, and also due to the increase of
asymmetry energy (for details see Ref.[5]).

To describe the quark phase an improved version of the MIT bag model is
used, in which the interactions between $u,~d,~s$ quarks inside the bag are
taken in a one-gluon exchange approximation [6]. For quark masses we choose $%
m_{u} = 5~MeV$, $m_{d} = 7~MeV$ and $m_{s} = 150~MeV$, and $\alpha_{s}=0.5$
for the strong interaction constant. Calculations are carried out for
different values of the bag parameter $B$ in the range of $60 \le B \le
120~MeV/fm^3$.

\begin{table}[h]
\caption{The Maxwellian phase transition parameters with and without $%
\protect\delta$-meson field.}\centering  
\begin{tabular}{|c|c|c|c|c|c|c|c|}
\hline
$B$ &  & $n_{N}$ & $n_{Q}$ & $P_{0}$ & $\varepsilon_{N}$ & $\varepsilon_{Q}$
& $\lambda$ \\
$MeV/fm^{3}$ &  & $fm^{-3}$ & $fm^{-3}$ & $MeV/fm^{3}$ & $MeV/fm^{3}$ & $%
MeV/fm^{3}$ &  \\ \hline
60 & $\sigma\omega\rho$ & 0.1220 & 0.2826 & 1.965 & 115.8 & 270.9 & 2.299 \\
\hline
60 & $\sigma\omega\rho\delta$ & 0.1207 & 0.2831 & 2.11 & 114.5 & 271.4 &
2.327 \\ \hline
69.3 & $\sigma\omega\rho$ & 0.246 & 0.3557 & 15.57 & 239.6 & 353.4 & 1.385
\\ \hline
69.3 & $\sigma\omega\rho\delta$ & 0.2241 & 0.3504 & 13.84 & 217.5 & 347.9 &
1.504 \\ \hline
80 & $\sigma\omega\rho$ & 0.3792 & 0.4819 & 48.54 & 384.5 & 501.8 & 1.159 \\
\hline
80 & $\sigma\omega\rho\delta$ & 0.3276 & 0.4525 & 37.95 & 328.8 & 468.6 &
1.278 \\ \hline
100 & $\sigma\omega\rho$ & 0.5506 & 0.7175 & 121.3 & 593.4 & 810 & 1.133 \\
\hline
100 & $\sigma\omega\rho\delta$ & 0.4746 & 0.6497 & 93.30 & 503.3 & 723.5 &
1.213 \\ \hline
120 & $\sigma\omega\rho$ & 0.6512 & 0.8975 & 180.2 & 728.9 & 1073 & 1.18 \\
\hline
120 & $\sigma\omega\rho\delta$ & 0.5729 & 0.8165 & 143.9 & 631.7 & 961.4 &
1.24 \\ \hline
\end{tabular}%
\end{table}
We assumed that deconfinement phase transition of nuclear matter into quark
matter is a usual first-order phase transition described by Maxwells
construction. The Table 2 represents the values of parameters of such phase
transition for different values of the bag parameter $B$ with ($%
\sigma\omega\rho\delta$) and without ($\sigma\omega\rho$) the $\delta$
-field. In this Table $n_{N}$ and $n_{Q}$ are the baryon number densities at
transition point of nuclear and quark phases, respectively, $\varepsilon_{N}$
and $\varepsilon_{Q}$ are the energy densities, $P_{0}$ is the phase
transition pressure, and $\lambda=\varepsilon_{Q}/(\varepsilon_{N}+%
\varepsilon_{Q})$ is the density jump parameter.

\section{Summary}

The presence of the $\delta$-meson field leads to the decrease of both
transition pressure, $P_{0}$, and baryon number densities $n_{N}$ and $n_{Q}$
. Besides, the jump parameter, $\lambda$ , is increased. According to
Ref.[7], in case of $\lambda>\lambda_{cr}=3/2$, the infinitisimal core of
the new phase is unstable. Our analysis show that in case of $B<B_{cr}=69.3
MeV/fm^3$, the density jump parameter satisfies the condition $%
\lambda>\lambda_{cr}$ , and the neutron star configurations with
infinetisimal quark cores are unstable.

\textbf{Acknowledgements.} This work is supported by the Ministry
of Education and Science of Armenia under project No.130.

\end{document}